\documentclass{appolb}
\usepackage{graphicx}
\begin{document}

\title{Eta-Mesic Nucleus: A New Form of Nuclear Matter %
\thanks{Talk presented by Q.H. at the International
Symposium on Meson Physics, held 1-4 October, 2008 at Krak\'ow, Poland.
All correspondences should be sent to haider@fordham.edu}}
\author{ Q. Haider
\address{Physics Department, Fordham University,
Bronx, N. Y. 10458}
\and
 L.C. Liu
\address{Theoretical Division, Los Alamos National Laboratory,
Los Alamos, N. M 87545}}
\date{\today}
\maketitle

\begin{abstract}
Formation of $\eta$-mesic nucleus, a bound state of an $\eta$ meson in a nucleus,
is reviewed in this paper. Three different theoretical approaches are used to
calculate the binding energies and widths of such nuclei. The effect of $\eta$-mesic
nucleus in pion double-charge-exchange reaction is discussed. Experimental
efforts by different groups to detect the nucleus are also discussed. The ramifications
of the theoretical and experimental studies of the bound state of $\eta$ in a nucleus
are pointed out.
\end{abstract}

\section{Introduction}\label{sec:1}
In the past, bound systems of strongly interacting particles in a nucleus or atom,
such as hyperons and mesons, have provided valuable information about various
aspects of hadron-nucleon interaction in a many-body environment. For example,
experimental and theoretical studies of hypernuclei led to
impressive advances in our knowledge of the $\Lambda N$ and $\Sigma N$ interactions.

Until recently it was believed that the
$\eta$ meson plays almost no role in nuclear physics because the $\eta NN$ coupling
constant is very small compared to $\pi NN$ and $\pi N\Delta$ coupling constants. This is in
sharp contrast to $\pi$-nucleus interaction which has been studied extensively. The
situation has, however, changed recently in medium- and high-energy nuclear reactions where
significant amount of pion induced $\eta$ production has been observed at pion energies
near 500~MeV~\cite{peng}. In view of this, there has been a surge of interest in
studying the $\eta$ meson within the framework of nuclear physics. Some of the studies
have led to interesting surprises, such as the prediction of the existence of a bound state
of $\eta$ meson in a nucleus, termed $\eta$-mesic nucleus, and the phenomenon of ``mesonic
compound nucleus'' in high energy pion double-charge-exchange reactions. In this paper, we
will describe the formation of the $\eta$-mesic nucleus and point out the relevant
physics that we can hope to learn by using the $\eta$ meson as a nuclear probe.

\section{Formation of Eta-Mesic Nucleus}\label{sec:2}

The existence of $\eta$-mesic nucleus was first predicted by us in 1986~\cite{hai1}.
It is a consequence of the attractive
interaction between the $\eta$ meson and all the nucleons in the nucleus. The
attractive nature of the interaction follows from the
work of Bhalerao and Liu~\cite{bhal} who found, from a detailed
coupled-channel analysis of $\pi N\rightarrow \pi N$,
$\pi N\rightarrow \pi\pi N$, and $\pi N\rightarrow \eta N$ reactions,
that near-threshold $\eta N$ interaction is attractive.

The binding energy $\epsilon_{\eta}$ and width
$\Gamma_{\eta}$ of the $\eta$-mesic nucleus are calculated by solving
the momentum-space relativistic three-dimensional
integral equation
\begin{equation}
\frac{{\bf k'}^{2}}{2\mu}\;{\psi}({\bf k'})
+ \int\;d{\bf k}<{\bf k}'\mid {V}\mid{\bf k}>{\psi}({\bf k})
=E{\psi}({\bf k}')\ ,
\label{eq:2.1}
\end{equation}
using the inverse-iteration method of Kwon and Tabakin~\cite{kwon}.
Here $<{\bf k'}\mid V \mid {\bf k}>$ are
momentum-space matrix elements of the $\eta$-nucleus
optical potential $V$, with
${\bf k}$ and ${\bf k}'$ denoting, respectively,
the initial and final $\eta$-nucleus relative momenta.
The $\mu$ is the reduced mass of the $\eta$-nucleus system and
$E=\epsilon_{\eta}+i\Gamma_{\eta}/2 \equiv
\kappa^{2}/2\mu$ is the complex eigenenergy.
For bound states, both $\epsilon_{\eta}$ and $\Gamma_{\eta}$ are negative.
Three different theoretical approaches to $V$ are used to calculate $E$,
and they are described below.

\subsection{Covariant $\eta$-Nucleus Optical Potential}\label{sec:2.1}

The first-order microscopic $\eta$-nucleus optical potential, after using
the covariant reduction scheme of Celenza et al.~\cite{cel1},
has the form\cite{hai1,hai4}
\begin{eqnarray}
 <{\bf k}'\mid V\mid{\bf k}> & = & \sum_{j}\int d{\bf Q}
<{\bf k}',-({\bf k}'+{\bf Q})\mid t(\sqrt{s_{j}})_{\eta N\rightarrow\eta N}
\mid \nonumber \\
 & \times & \mid {\bf k}, -({\bf k}+{\bf Q})> \phi^{*}_{j}(-{\bf k}'-{\bf Q})
\phi_{j}(-{\bf k}-{\bf Q})\ ,
\label{eq:2.1.6}
\end{eqnarray}
where the off-shell $\eta N$ interaction
$t_{\eta N\rightarrow\eta N}$ is weighted by the product of the nuclear
wave functions $\phi^{*}_{j}\phi_{j}$ corresponding to having the
nucleon $j$ at the momenta $-({\bf k}+{\bf Q})$ and $-({\bf k}'+{\bf Q})$
before and after the collision, respectively. The $\eta$-nucleus interaction
$V$ is related to the elementary $\eta N$ process by the kinematical
transformations of Liu and Shakin~\cite{liu1}. The
$\eta N$ invariant mass $\sqrt{s_{j}}$ in the
c.m. frame of the $\eta$ and the nucleon $j$ is given by
\begin{eqnarray}
s_{j} & = & [\{ W-E_{C,j}({\bf Q})\}^{2}-{\bf Q}^{2}] \nonumber \\
 & \simeq &
\left [ M_{\eta}+M_{N}-\mid\epsilon_{j}\mid \; -\; \frac{{\bf Q}^{2}}
{2M_{C,j}}\;\left ( \frac{M_{\eta}+M_{A}}{M_{\eta}+M_{N}} \right )
\right ]^{2} \nonumber \\
& < & (M_{\eta}+M_{N})^{2}\ ,
\label{eq:2.1.7}
\end{eqnarray}
where ${\bf Q}$,  $E_{C,j}$ and $M_{C,j}$ are, respectively, the
momentum, total energy, and mass
of the core nucleus arising from removing a nucleon $j$ of
momentum $-({\bf k}+{\bf Q})$ and binding energy $\mid\epsilon_{j}\mid$
from the target nucleus having the momentum $-{\bf k}$.
Calculation of $V$ involves full off-shell kinematics and integration over the
Fermi motion variable ${\bf Q}$. It requires knowledge of the basic
$t_{\eta N\rightarrow\eta N}$ at subthreshold energies~\cite{bhal}. All
kinematic quantities are calculated using well-established Lorentz transformations.
For near threshold $\eta N$ interaction, only one resonance of the $N^{*}$ isobar has
to be considered for each partial wave. They are $S_{11}, \; P_{11},$
and $D_{13}$ resonances. The $\eta N$ interaction parameters are taken
from ref.\cite{bhal}. Details of the calculation can be found in refs.~\cite{hai1,hai4}.

\begin{table}
\caption{Binding energies and half-widths (both in MeV) of $\eta$-mesic
nuclei given by the full off-shell calculation. No bound state
solutions were found for $A \leq 10$. }
\label{table:1}
\begin{center}
\begin{tabular}{ccc} \hline\hline
Nucleus & Orbital ($n\ell$) & $\epsilon_{\eta}+ i\Gamma_{\eta}/2$ \\ \hline\hline
$^{12}$C  &  $1s$ & $ -(1.19 + 3.67i)$   \\
$^{16}$O  &  $1s$ & $ -(3.45 + 5.38i)$   \\
$^{26}$Mg &  $1s$ & $ -(6.39 + 6.60i)$   \\
$^{40}$Ca &  $1s$ & $ -(8.91 + 6.80i)$   \\
$^{90}$Zr &  $1s$ & $-(14.80 + 8.87i)$   \\
          &  $1p$ & $ -(4.75 + 6.70i)$   \\
$^{208}$Pb&  $1s$ & $-(18.46 + 10.11i)$  \\
          &  $2s$ & $ -(2.37 + 5.82i)$   \\
          &  $1p$ & $-(12.28 + 9.28i)$   \\
          &  $1d$ & $ -(3.99 + 6.90i)$   \\ \hline\hline
\end{tabular}
\end{center}
\end{table}

The results of the calculation are presented in table~\ref{table:1}. The nuclear
wave functions in eq.(\ref{eq:2.1.6}) are derived from experimental form factors with
the proton finite size corrected for.
From the table 1, it can be seen that $\eta$ can be bound in nuclei with
mass number $A>10$. The $\eta$-nucleus interaction is not strong enough to
have a bound state in lighter nuclei. This is because there is a reduction in
the strength of the $\eta$-nucleus interaction at subthreshold energies. The
number of nuclear orbitals in which $\eta$ can be bound increases with increasing
mass number.

It should be mentioned that the calculations do not include the
effects of Pauli blocking. However, estimates of the blocking using local-density
approximation show about 5\% reduction in the widths~\cite{hai1}. The calculated
widths of $\eta$-mesic nuclei vary from 10 to 13~MeV in lighter nuclei and are
between 15 and 20~MeV for the heavier ones. About 95\% of the width is due to the
decay of the $\eta$-mesic nucleus with the emission of one pion and one nucleon.
The remaining 5\% is due to the emission of two pions~\cite{hai5}. As absorptions of
$\eta N\rightarrow N$ and $\eta NN\rightarrow NN$ are strongly suppressed kinematically,
their contributions to the width are insignificant.
Widths of the order of keV may be possible if $\eta NNN\rightarrow NNN$ can take place.
In this latter case, the $\eta$ can then share its energy and momentum with three nucleons.

The binding energy is very sensitive to the $\eta N$ interaction coupling constant
$g_{_{\eta NN^{*}}}$. The value of $g_{_{\eta NN^{*}}}$ determined in ref.~\cite{bhal} is 0.77.
This leads to a value of 0.28~fm for the real part of the $s$-wave $\eta N$ scattering length
$a_{\eta N}$, corresponding to an attractive interaction. As an example, it has been shown by
Haider and Liu~\cite{hai6} that the bound state of $\eta$ in $^{15}$O increases
with the coupling constant.
For $g_{_{\eta NN^{*}}} > 0.90$, however, the binding energy starts to decrease, and then
ceases to exist when it becomes greater than 1.0. Furthermore, $\Re ({a_{\eta N}})$ also decreases
for $g_{_{\eta NN^{*}}} > 0.85$ and becomes repulsive at large values.

The existence of at least one bound state of $\eta$ in medium mass nuclei
of radius $R=r_{0}A^{1/3}$ can be understood by considering an
equivalent square-well (complex) potential of depth $V_{0}=U_{0}+iW_{0}$.
One $s-$wave bound state is possible if the condition
\begin{equation}
\frac{\pi^{2}}{8\mu } < (\mid U_{0}\mid R^{2}) < \frac{9\pi^{2}}{8\mu}
\label{eq:2.1.8}
\end{equation}
is satisfied~\cite{schi}.
In terms of the $s$-wave $\eta N$ scattering length $a_{_{\eta N}}$, the condition is
\begin{equation}
X < \Re{(a_{_{\eta N}})} < 9X, \;\;\; X=\frac{\pi^{2}r_{0}}{12A^{2/3}}\left (
1+\frac{M_{\eta}}{M_{N}} \right )^{-1},
\label{eq:2.1.9}
\end{equation}
and the potential is given by
\begin{equation}
V_{0} = -197.3 \left ( \frac{3a_{_{\eta N}}}{2r_{0}^{3}} \right ) \left ( 1+\frac{M_{\eta}}{M_{N}}
\right ) \left ( \frac{M_{\eta}+M_{A}}{M_{\eta}M_{A}} \right).
\label{eq:2.1.10}
\end{equation}

\begin{table}
\caption{Parameters $X$ and $9X$ (both in fm) of the equivalent square-well potential
calculated with $a_{_{\eta N}}=(0.28+0.19i)$ fm.}
\label{table:2}
\begin{center}
\begin{tabular}{ccc} \hline\hline
Nucleus &  $X$  & $9X$ \\ \hline\hline
$^{12}$C &  0.109 & 0.981 \\
$^{16}$O &  0.090 & 0.810 \\
$^{26}$Mg & 0.065 & 0.585 \\
$^{40}$Ca & 0.048 & 0.432 \\
$^{90}$Zr & 0.029 & 0.261 \\
$^{208}$Pb & 0.016 & 0.144 \\ \hline\hline
\end{tabular}
\end{center}
\end{table}
\noindent
In the above expression, the masses are in fm$^{-1}$ and $V_{0}$ is in MeV.
The values of $X$ and $9X$ are given in table~\ref{table:2} for
$r_{0}=1.1$~fm and $a_{_{\eta N}} = (0.28 + 0.19i)$~fm \cite{bhal}. For nuclei
with $A \leq 10$, one has to use the actual value of $R$ which is substantially larger
than the value given by $1.1A^{1/3}$~fm. Additionally, the use of equivalent
square-well potential and the omission of the imaginary part of the potential
break down for lighter nuclei. Consequently, eqs.(\ref{eq:2.1.8})-(\ref{eq:2.1.10})
should not be applied to nuclei with $A \leq 10$. Instead,
detailed calculations should be performed. In Zr and Pb, $\Re (a_{_{\eta N}}) > 9X.$
This explains why there are more than  one bound state in these nuclei.

\subsection{Factorization Approximation}\label{sec:2.2}
In the factorization approximation (FA), the $\eta$N
scattering amplitude in eq.(\ref{eq:2.1.6}) is taken out of the ${\bf Q}$
integration and evaluated at a fixed momentum $<{\bf Q}>$ given by
\begin{equation}
<{\bf Q}> = -\;\left ( \frac{A-1}{2A} \right )\;({\bf k}'-{\bf k})\ .
\label{eq:2.2.1}
\end{equation}
This choice of $<Q>$ corresponds to a motionless target nucleon fixed before
and after the $\eta N$ interaction. It has the virtue of preserving the
symmetry of the $t$-matrix
with respect to the interchange of ${\bf k}$ and ${\bf k}'$. With this approximation,
the $\eta$-nucleus potential can be written as

\begin{eqnarray}
<{\bf k}'\mid V_{FA}\mid {\bf k}> & =  &<{\bf k}', -({\bf k}'+<{\bf Q}>) \mid
t(\sqrt{\overline{s}})_{\eta N\rightarrow\eta N} \mid \nonumber \\
 & \times & \mid {\bf k},-({\bf k}+<{\bf Q}>)>
f({\bf k}'-{\bf k})\ ,
\label{eq:2.2.2}
\end{eqnarray}
where
\begin{equation}
f({\bf k}'-{\bf k}) = \sum_{j}\int d{\bf Q}\;\phi_{j}^{*}(-{\bf k}'-{\bf Q})
\phi_{j}(-{\bf k}-{\bf Q})\ ,
\label{eq:2.2.3}
\end{equation}
is the nuclear form factor having the normalization $f(0)=A$.
Guided by the expression for $\sqrt{s_{j}}$, the $t$-matrix in eq.(\ref{eq:2.2.2})
is evaluated at
$\sqrt{\overline{s}} =M_{\eta}+M_{N}-\Delta \equiv \sqrt{s_{th}} -
\Delta$, with $\Delta$ being an energy shift parameter. A downward shift
of $\sim 30$~MeV is used to fit $\pi$N scattering data~\cite{liu1}.

The bound-state solutions obtained from using the covariant factorized
potential with $\Delta =0, \;10, \; 20, \;30$~MeV are presented in
table~\ref{table:3}. The interaction parameters used in the FA calculations
are same as those used for the off-shell calculations. The nuclear form factors
used in the calculations can be found in refs.\cite{hofs,jage}. A comparison
between tables~\ref{table:1} and \ref{table:3} indicates that the FA results
with $\Delta = 30$~MeV are close to the off-shell results. This indicates that
the $\eta N$ interaction in $\eta$ bound-state formation takes place at energies
30~MeV below the free-space threshold.

\begin{table}
\caption{Binding energies and half-widths (both in MeV) of $\eta$-mesic
nuclei obtained with the factorization approach for different
values of the energy shift parameter $\Delta$ (in MeV).}
\label{table:3}
\begin{center}
\footnotesize
\begin{tabular}{cccccc} \hline\hline
Nucleus & Orbital ($n\ell$) & $\Delta = 0$ & $\Delta = 10$ & $\Delta = 20$
& $\Delta = 30$ \\ \hline\hline
$^{12}$C  &  $1s$ & $ -(2.18 + 9.96i) $  &  $ -(1.80 + 6.80i)$
& $-(1.42 + 5.19i)$ & $-(1.10 + 4.10i)$ \\
$^{16}$O  &  $1s$ & $ -(4.61 + 11.57i)$  &  $ -(3.92 + 8.13i)$
& $-(3.33 + 6.37i)$ & $-(2.84 + 5.17i)$ \\
$^{20}$Ne & $1s$ &$-(6.52+12.86i)$ & $-(5.63+9.16i)$ & $-(4.90+7.26i)$ &
$-(4.29+5.96i)$ \\
$^{24}$Mg & $1s$ &$-(9.26+14.90i)$ & $-(8.09+10.75i)$ & $-(7.15+8.60i)$&
$-(6.37+7.13i)$ \\
$^{26}$Mg &  $1s$ & $ -(10.21 + 15.41i)$  &  $ -(8.95 + 11.17i)$
& $-(7.94 + 8.97i)$ & $-(7.11 + 7.46i)$ \\
$^{28}$Si & $1s$ & $-(10.84+15.70i)$ & $-(9.53+11.40i)$ & $-(8.49+9.18i)$ &
$-(7.62+7.65i)$ \\
$^{32}$S & $1s$ & $-(11.94+16.18i)$ & $-(10.56+11.80i)$ & $-(9.47+9.53i)$ &
$-(8.55+7.97i)$ \\
$^{40}$Ca &  $1s$ & $ -(14.34 + 17.06i)$  &  $ -(12.75 + 12.55i)$
& $-(11.53 + 10.21i)$ & $-(10.51 + 8.59i)$ \\
$^{46}$Ti & $1s$ & $-(15.40+17.12i)$ & $-(13.73+12.66i)$ & $-(12.46+10.34i)$ &
$-(11.40+8.73i)$ \\
$^{52}$Cr & $1s$ & $-(16.42+16.99i)$ & $-(14.65+12.63i)$ & $-(13.33+10.35i)$ &
$-(12.24+8.77i)$ \\
& $1p$ & $-(2.10+15.09i)$ & $-(1.40+10.54i)$ & $-(0.75+8,21i)$ & $-(0.19+6.62i)$\\
$^{56}$Fe & $1s$ & $-(16.73+16.81i)$ & $-(14.94+12.53i)$ & $-(13.60+10.28i)$ &
$-(12.51+8.72i)$ \\
& $1p$ & $-(2.68+15.00i)$ & $-(2.00+10.53i)$ & $-( 1.34+8.25i)$ & $-(0.76+6.69i)$ \\
$^{58}$Ni & $1s$ & $-(17.04+16.88)$ & $-(15.23+12.59i)$ & $-(13.88+10.34i)$ &
$-(12.77+8.78i)$ \\
& $1p$ & $-(3.17+14.82i)$ & $-(2.43+10.47i)$ & $-(1.73+8.24i)$ & $-(1.13+6.71i)$ \\
$^{70}$Zn & $1s$ & $-(18.57+17.34i)$ & $-(16.62+12.98i)$ & $-(15.20+10.69i)$ &
$-(14.03+9.10i)$ \\
& $1p$ & $-(5.02+15.29i)$ & $-(4.20+10.94i)$ & $-(3.42+8.71i)$ & $-(2.75+7.18i)$ \\
$^{90}$Zr &  $1s$ & $-(21.32 + 18.59i)$  &  $-(19.15 + 13.97i)$
& $-(17.58 + 11.54i)$ & $-(16.29 + 9.84i)$ \\
&  $1p$ & $ -(8.27 + 16.01i)$  &  $ -(7.19 + 11.47i)$
& $-(6.23 + 9.48i)$ & $-(5.40 + 7.94i)$ \\
$^{208}$Pb&  $1s$ & $-(24.06 + 19.18i)$ &  $-(21.88 + 14.44i)$
          & $-(20.28 + 11.96)$ & $-(18.96 + 10.22i)$ \\
          &  $2s$ & $ -(4.89 + 11.04i)$  &  $ -(3.67 + 8.28i)$
          & $-(2.81 + 6.79i)$ & $-(2.12 + 5.72i)$ \\
          &  $1p$ & $-(18.33 + 18.97i)$  &  $ -(16.31 + 14.27i)$
          & $-(14.81 + 11.79i)$ & $-(13.56 + 10.06i)$ \\
          &  $1d$ & $ -(8.27 + 14.07i)$  &  $ -(6.17 + 10.56i)$
          & $-(5.58 + 8.71i)$ & $-(4.66 + 7.41i)$ \\ \hline\hline
\end{tabular}
\end{center}
\end{table}

\begin{figure}
\includegraphics[angle=270,width=1\columnwidth]{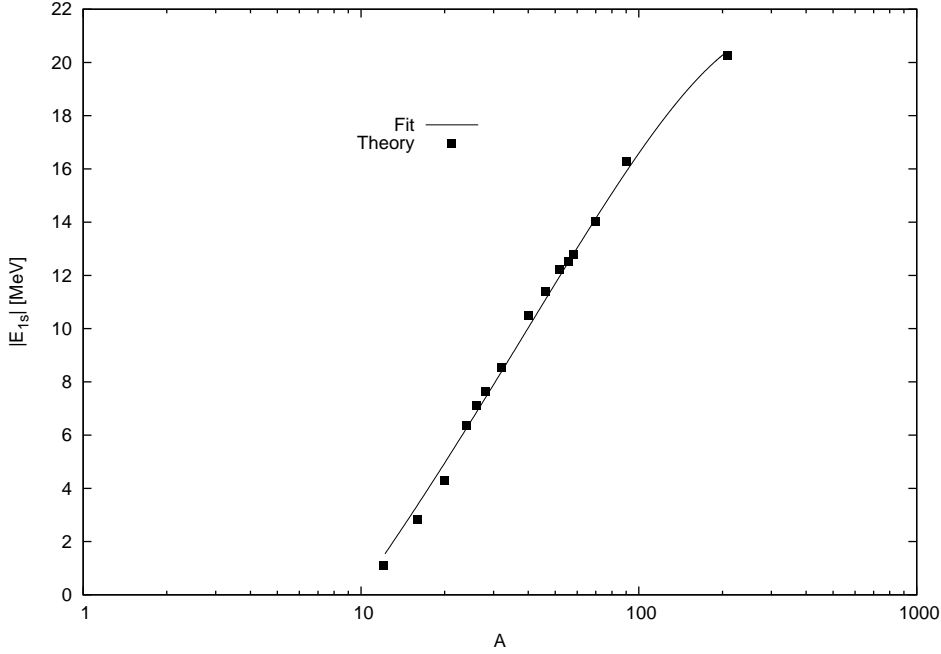}\caption[dependence]{Depenendence of the binding energy
of $\eta$-mesic nucleus on the mass number $A$.}
\label{fig1}
\end{figure}

The mass dependence of the $1s$ binding energies for the $\Delta = 30$~MeV case is
shown in figure \ref{fig1}. The binding energies have been fitted empirically
with the formula

\begin{equation}
\epsilon_{\eta}= aA+bA^{2/3}+cA^{1/3}+d, \;\;\;   12 \leq A \leq 208,
\label{eq:2.2.4}
\end{equation}
where $a=0.0003, \; b=-0.9562, \; c=13.06$, and $d=-23.46$ (all in MeV).
The above equation resembles the semi-empirical mass formula with $a$ as the volume energy
term and $b$ as the surface energy term.

\subsection{On-Shell Optical Potential}\label{sec:2.3}
First-order low energy $\eta$-nucleus on-shell optical potential is
\begin{equation}
<{\bf k'}\mid V_{ON}\mid {\bf k}> = -\frac{1}{4\pi^{2}\mu} \left ( 1+\frac{M_{\eta}}{M_{N}} \right )
a_{_{\eta N}} f({\bf k'}-{\bf k}).
\label{eq:2.3.1}
\end{equation}
As will be seen later, $V_{ON}$
corresponds to $V_{FA}$ with no energy shift $(\Delta = 0)$
and gives an upper limit to the value of $\epsilon_{\eta}$.
The only input for the on-shell calculation is the $s$-wave $\eta N$ scattering length $a_{_{\eta N}}$.

The scattering length is not directly measurable and its value is model dependent.
Different models predict different values of $a_{_{\eta N}}$~\cite{hai4}.
The value of $a_{_{\eta N}}$ in the literature varies from
$0.27 \leq \Re ({a_{_{\eta N}}}) \leq 1.05, \;\; 0.19 \leq \Im ({a_{_{\eta N}}}) \leq 0.37.$
A reason for this large range is unavailability of data on $\eta N$ elastic scattering, which
is essential in determining the value of $a_{_{\eta N}}$.

\normalsize
\begin{table}
\caption{Binding energies and half-widths (both in MeV) of $\eta$-mesic
nuclei ($1s$ state only) given by the on-shell optical potential
for two different values of the scattering length $a_{_{\eta N}}$.
No bound state exists in $^{3}$He.}
\label{table:4}
\begin{center}
\begin{tabular}{cccc} \hline\hline
Nucleus & F.A. ($\Delta = 0$) & $a_{_{\eta N}}=(0.28+0.19i)$~fm &
$a_{_{\eta N}}=(0.51+0.21i)$~fm \\ \hline\hline
$^{4}$He& $-$ & $-$  & $-(6.30+11.47i)$   \\
$^{6}$Li  & $-$  &             $-$    & $-(3.47+6.79i)$    \\
$^{9}$Be  & $-$  &            $-$     & $-(13.78+12.45i)$  \\
$^{10}$B  & $-$  & $-(0.93+8.70)$    & $-(15.85+13.05i)$  \\
$^{11}$B  & $-$  & $-(2.71+10.91i)$  & $-(20.78+15.42i)$  \\
$^{12}$C  &  $-(2.18 + 9.96i)$& $-(2.91+10.22i)$  & $-(19.61+14.20i)$  \\
$^{16}$O  &  $-(4.61 + 11.57i)$ & $-(5.42+11.43i)$  & $-(23.26+14.86i)$  \\
$^{20}$Ne & $-(6.52+12.86i)$ & $-(7.44+12.61i)$ & $-(26.72+15.94i)$ \\
$^{24}$Mg & $-(9.26+14.90i)$ & $-(10.34+14.40i)$ & $-(32.00+17.59i)$ \\
$^{26}$Mg &  $-(10.21+15.41i)$ & $-(11.24+14.76i)$ & $-(33.11+17.73i)$  \\
$^{28}$Si & $-(10.84+15.70i)$ & $-(11.90+15.05i)$ & $-(34.06+17.96i)$ \\
$^{32}$S & $-(11.94+16.18i)$ & $-(13.10+15.61i)$ & $-(35.90+18.46i)$ \\
$^{40}$Ca &  $-(14.34+17.06i)$ & $-(15.46+16.66i)$ & $-(38.85+19.16i)$  \\
$^{46}$Ti & $-(15.40+17.12i)$ & $-(16.50+17.07i)$ & $-(39.82+19.44i)$ \\
$^{52}$Cr & $-(16.42+16.99i)$ & $-(17.46+17.45i)$ & $-(40.5i+19.76i)$ \\
$^{56}$Fe & $-(16.73+16.81i)$ & $-(17.76+17.52i)$ & $-(40.67+19.85i)$ \\
$^{58}$Ni & $-(17.04+16.88i)$ & $-(18.09+17.67i)$ & $-(41.13+20.02i)$ \\
$^{70}$Zn & $-(18.57+17.34i)$ & $-(19.63+18.47i)$ & $-(43.52+20.93i)$ \\
$^{90}$Zr &  $-(21.32+18.59i)$ & $-(22.41+19.97i)$ & $-(48.40+22.60i)$  \\
$^{208}$Pb&  $-(24.06+19.18i)$ & $-(24.55+19.57i)$ & $-(50.27+21.42i)$  \\ \hline\hline
\end{tabular}
\end{center}
\end{table}

The binding energies and half-widths given by eq.(\ref{eq:2.3.1})
are presented in table~\ref{table:4} for two different
values of $a_{_{\eta N}}$.
The nuclear form factors are the same as those used in the FA calculations.
For these two scattering lengths, no bound state can exist in $^{3}$He.
Upon comparing the third column of table~\ref{table:4} with the off-shell
calculation (table~\ref{table:1}), it can be seen that
the on-shell approximation predicts
more strongly bound $\eta$-mesic nuclei.
Also, as expected, the on-shell
results for $a_{_{\eta N}}=(0.28+0.19i)$~fm are similar to those of FA with
$\Delta =0$~MeV.

\section{Eta-Mesic Nucleus and Pion DCX}\label{sec:3}
Eta-mesic nucleus can affect high-energy pion double-charge-exchange
(DCX) reactions through the formation of ``mesonic compound nucleus''~\cite{hai3}.
For pion kinetic energies $T_{\pi} > 400$~MeV, $\eta$ production channel is open in most nuclei.
As a result, the DCX reaction can proceed via $\pi^{+}\rightarrow \pi^{0}\rightarrow \pi^{-}$
or $\pi^{+} \rightarrow\eta\rightarrow\pi^{-}$.
While $\pi^{0}$ is in the continuum, $\eta$ can either be in the continuum or in a nuclear
bound state. The DCX amplitudes associated with the $\eta$-nucleus bound states, on the
other hand, have resonance structure.

\begin{figure}
\includegraphics[width=1\columnwidth]{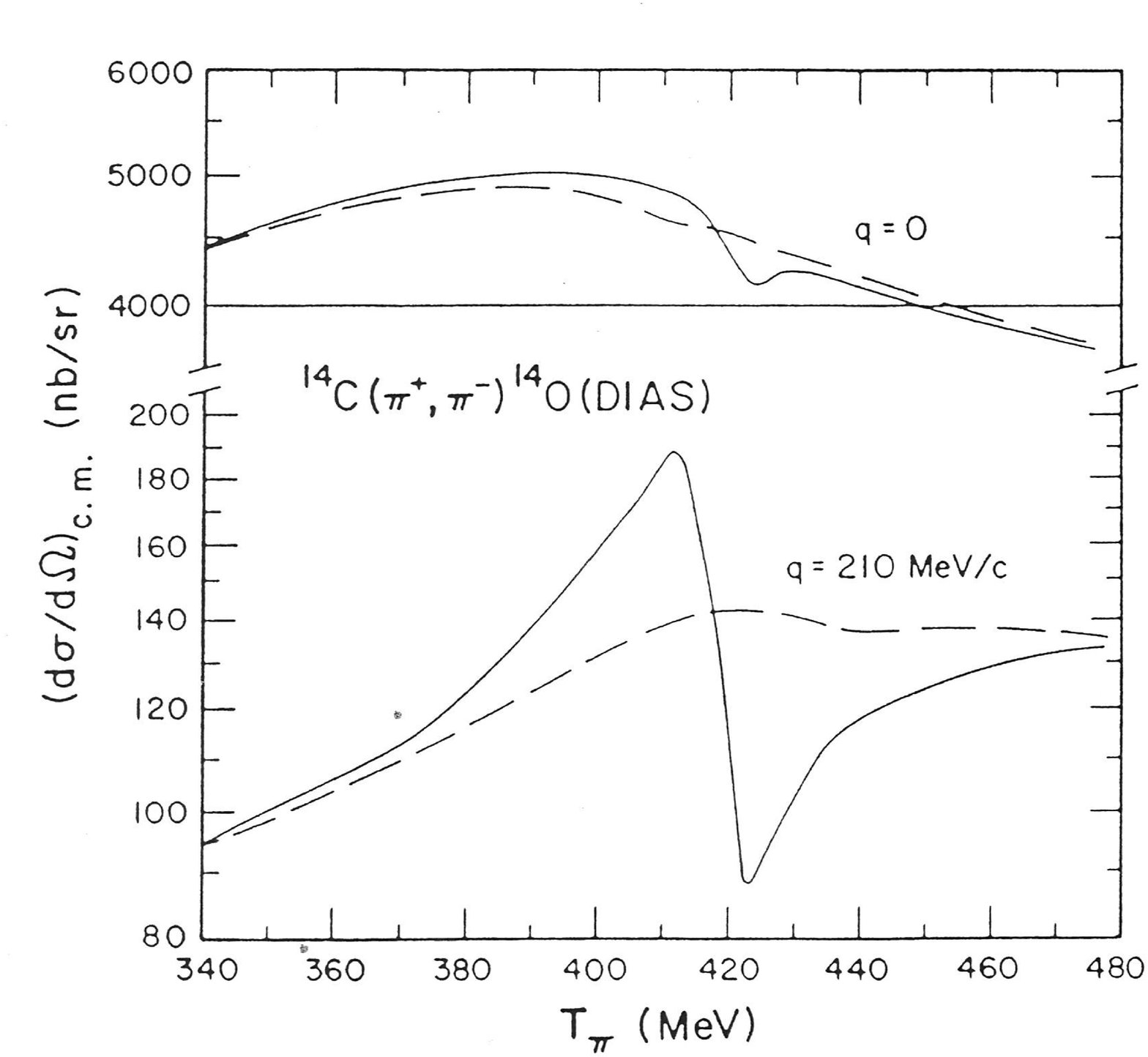}\caption[dependence]{The calculated energy dependences of
the DCX reaction described in the text leading to the double isobaric analog state.}
\label{fig2}
\end{figure}

The calculated cross sections for the reaction $^{14}$C$(\pi^{+},\pi^{-})^{14}$N
as a function of $T_{\pi}$ at momentum transfers $q=0$ and 210~MeV/c
are shown in figure \ref{fig2}. The solid curves represent the contribution from the
resonant amplitude associated with the formation of $\eta$-mesic nucleus. The dashed
curves represent contribution from the nonresonant amplitude only. The interference
of these amplitudes is responsible for the presence of narrow resonance structure at
$T_{\pi}\sim 415$~MeV for $q=210$~MeV/c case. The width of the resonance is about 10~MeV, which reflects
the width ($\sim 11$~MeV) of the $\eta$-mesic nucleus used in the calculations. The DCX
studies, therefore, can be used as an alternative way to determine the width of $\eta$-mesic
nucleus. Other processes in which one can expect to see these kind of effects is
$(\pi ,\pi ')$ reactions leading to certain specific final states. The $\eta$-nucleus
bound state amplitude for these kind of reactions is not small, in comparison to the
nonresonant amplitude. The study of the resonance pattern in the energy dependence of
the cross section can yield information on the relative phase between resonant
and nonresonant amplitudes.

\section{Experimental Search for Eta-Mesic Nucleus}\label{sec:4}
Several experiments were performed to detect $\eta$-mesic nucleus. The first
experiment at Brookhaven National Laboratory~\cite{chri} in 1987 was based
on the work of Liu and Haider~\cite{hai2}. A second experiment, based on the pion
DCX calculations~\cite{hai3} was done at LAMPF by Johnson et al.~\cite{john}.
While these experiments could not confirm unambiguously the existence of
$\eta$-mesic nucleus, they did not rule out such a possibility either.
In recent publications, Sokol et al.~\cite{sok1} claim to have observed
$\eta$-mesic nucleus in experiments involving photo-mesonic reactions.

Recently, an experiment has been done at J\"{u}lich by the COSY-GEM
Collaboration~\cite{cosy}, making use of the transfer reaction
p$+^{27}$Al$\rightarrow^{3}$He$+^{25}$Mg$_{\eta}$. Analysis of
the data clearly indicates the detection of bound state of $\eta$ in $^{25}$Mg.
Vigorous efforts are also underway to detect bound state of $\eta$ in
$^{3,4}$He. It is claimed that  a ``quasibound state'' of $\eta$ in
$^{3}$He may have been observed in the
$dp\rightarrow^{3}$He$\eta$ reaction~\cite{mers,coll}. This is in sharp contrast
to the findings of the COSY-11 and COSY-at-WASA groups~\cite{wasa}, also looking
for the bound state of $\eta$ in $^{3}$He in $dp$ collisions. Within the
statistical sensitivity achieved by them, they could not confirm the existence
of $\eta - ^{3}$He bound state. This is in agreement with the predictions presented
in this paper.

\section{Conclusions} \label{sec:5}
The formation of $\eta$-mesic nucleus is a natural consequence of the attraction
between the $\eta$ meson and the nucleon at very low energies. However, the
attraction is not strong enough to bind an $\eta$ onto a single nucleon. The
bound state formation is possible in a finite nuclei with mass number greater than 10.

The calculated binding energies and widths of $\eta$-nucleus
bound states strongly depend on the subthreshold dynamics of the $\eta N$
interaction. The present analysis
indicates that the average $\eta N$ interaction energy in mesic-nucleus
formation is below the threshold. What matters for the bound-state formation
is not the $\eta N$ interaction at the threshold but the effective in-medium
interaction. Because the subthreshold behavior of
$\eta N$ interaction is very model dependent, it is useful
for theorists to publish not only the $\eta$-nucleon scattering length
$a_{_{\eta N}}$, but also the corresponding subthreshold values
as a function of the shift parameter $\Delta$.

The downward shift in the effective interaction energy
can lead to a substantial reduction of the attraction
of in-medium $\eta$-nucleon interaction with respect to its
free-space value. Consequently, predictions based upon using free-space
$\eta N$ scattering length inevitably overestimate
the binding of $\eta$. One must bear this in mind when using the
predictions given by such calculations
as guide in searching for $\eta$-nucleus bound states.

Recent experimental confirmation~\cite{cosy} of the existence of
$\eta$-mesic nucleus will enable us to improve upon the existing models
or identify additional physics that has to be incorporated in them.
Because the binding energies of $\eta$ meson
depend strongly on the coupling between the $\eta N$ and the $N^{*}(1535)$
channels~\cite{bhal}, studies of $\eta$-mesic nucleus will be able
to yield detailed information on the $\eta NN^{*}$  coupling constant
involving bound nucleons. It can also lead to a new class of nuclear phenomenon,
$\eta$-mesic compound nucleus resonances. An awareness of this phenomenon
could be helpful to the analysis of nuclear reactions at energies above
the threshold for $\eta$ production.

The $\eta$-mesic nuclear levels correspond to an excitation energy of
$\sim 540$ MeV, to be compared with an excitation energy of
$\sim 200$ MeV associated with the $\Lambda$- and $\Sigma$-hypernuclei.
The existence of nuclear bound states with such high excitation
energies provides the possibility of
studying nuclear structure far from equilibrium.



\begin{thebibliography}{5}
 \bibitem{peng} J.C. Peng, {\it Hadronic Probes and Nuclear Interactions},
 AIP Conf. Proc. No. {\bf 133}, 255 (1985).
 \bibitem{hai1} Q. Haider and L.C. Liu, {\it Phys. Lett.} {\bf B 172}, 257
 (1986); {\bf B 174}, 465E (1986).
 \bibitem{bhal}   R.S. Bhalerao and L.C. Liu, {\it Phys. Rev. Lett}.
 {\bf 54}, 865 (1985).
 \bibitem{kwon} Y.R. Kwon and F.B. Tabakin, {\it Phys. Rev}. {\bf C 18},
 932 (1978).
 \bibitem{cel1} L.S. Celenza, M.K. Liou, L.C. Liu, and C.M. Shakin,
  {\it Phys. Rev}. {\bf C 10}, 398 (1974).
 \bibitem{hai4} Q. Haider and L.C. Liu, {\it Phys. Rev}. {\bf C 66}, 0425208 (2002).
 \bibitem{liu1} L.C. Liu and C.M. Shakin, {\it Prog. Part. and Nucl. Phys}.
 {\bf 5}, 207 (1980).
 \bibitem{hai5} Q. Haider and L.C. Liu, {\it Intersections Between Particle and Nuclear
 Physics}, AIP Conf. Proc. No. {\bf 150}, 930 (1986).
 \bibitem{hai6} Q. Haider and L.C. Liu, {\it Phys. Lett.} {\bf B 209}, 11 (1988).
 \bibitem{schi} L.I. Schiff, {\it Quaantum Mechanics} (3rd ed., Mc-Graw Hill, N.Y., 1968).
 \bibitem{hofs} H.R. Collard, L.R.B. Elton, and R. Hofstader,
  {\it Numerical Data and Functional Relationships in Science and
  Technology,} Vol.~2 Nuclear Radii, edited by H. Schopper
  (Springer-Verlag, N.Y., 1967).
  \bibitem{jage} C.W. de Jager, H. de Vries, and C. de Vries,
 {\it Atomic Data and Nuclear Data Tables} {\bf 14}, 479 (1974).
 \bibitem{hai3} Q. Haider and L.C. Liu, {\it Phys. Rev}. {\bf C 36},
 1636 (1987).
 \bibitem{chri} R.E. Chrien et al., {\it Phys. Rev. Lett}. {\bf 60},
  2595 (1988).
  \bibitem{hai2} L.C. Liu and Q. Haider, {\it Phys. Rev}. {\bf C 34},
 1845 (1986).
  \bibitem{john} J.D. Johnson et al., {\it Phys. Rev}. {\bf  C 47},
  2571 (1993).
   \bibitem{sok1} G.A. Sokol et al., {\it Particles and Nuclei Letters}
    {\bf 5}, 71 (2000); see also {\it Proceedings of the IX International
    Seminar on Electromagnetic
    Interaction of Nuclei at Low and Medium Energies,} Moscow, 2000, p.214.
 \bibitem{cosy} A. Budzanowski et al., {\it Phys. Rev.} {\bf C79}, 012201(R) (2009).
 \bibitem{mers} T. Mersmann et al., {\it Phys. Rev. Lett}. {\bf 98}, 242301 (2007).
 \bibitem{coll} C. Wilkin et al., {\it Phys. Lett}. {\bf B 654}, 92 (2007).
 \bibitem{wasa} J. Smyrski et al., {\it Phys. Lett}. {\bf B 649}, 258 (2007);
 {\it Nucl. Phys}. {\bf A 790}, 438 (2007); {\it Acta Physica Slovenia}
 {\bf 56}, 213 (2006).
 \end{thebibliography}
\end{document}